\documentclass[a4paper,fleqn,usenatbib]{mnras}

\usepackage[T1]{fontenc}
\usepackage{ae,aecompl}

\usepackage{graphicx}	
\usepackage{amsmath}	
\usepackage{amssymb}	

\title[A distortion solution for the Bok telescope with four CCD chips]{A distortion solution for the Bok telescope with four CCD chips}

\author[N. Wang et al.]{
N. Wang,$^{1,3}$
Q. Y. Peng,$^{1,3}$\thanks{E-mail: tpengqy@jnu.edu.cn (KTS)}
X. Zhou,$^{2,3}$
X. Y. Peng,$^{2,3}$
H. W. Peng$^{1,3}$
\\
$^{1}$Department of Computer Science, Jinan University, Guangzhou 510632, China\\
$^{2}$Key Laboratory of Optical Astronomy, National Astronomical Observatories, Chinese Academy of Sciences, Beijing 100012, China\\
$^{3}$Sino-French Joint Laboratory for Astrometry, Dynamics and Space Science, Jinan University, Guangzhou 510632, China\\
}

\date{Received ${\times}$ ${\times}$, 2018; accepted ${\times}$ ${\times}$, 2018}

\pubyear{2018}

\begin{document}
\label{firstpage}
\pagerange{\pageref{firstpage}--\pageref{lastpage}}
\maketitle

\begin{abstract}
The Beijing-Arizona Sky Survey (BASS) is an imaging survey and uses the 2.3 m Bok telescope at Kitt Peak. In order to tap the astrometric potential of the Bok telescope and improve the astrometry of BASS, a distortion solution for the Bok telescope is made. In the past, we used a single lookup table to correct all the positional errors. However, this method can not be applied to the reduction of the observation for the Bok telescope where four CCD chips are equipped with. Then quite different from our previous method, two third-order polynomials were used to fit the lookup table. By using the polynomial Geometric Distortion (called GD hereafter) correction the astrometry of BASS is improved greatly. Moreover, an additional lookup table correction is found to be more effective to obtain a final GD. The results show that the positional measurement precision of the appropriate bright stars is estimated at about 20 mas and even better in each direction. Besides, the relative positions of the chips are measured. The change of the inter-chip gaps in horizontal or vertical is no more than 5 pixel between 2016 and 2017 and the change of the roll angle is no more than 0.1 degree.
\end{abstract}




\section{Introduction}
The Beijing-Arizona Sky Survey (BASS) is a new g- and r-band imaging survey conducted by National Astronomical Observatory of China (NAOC) and Steward Observatory \citep{zou+etal+2017+aj}. BASS uses the 2.3 m Bok telescope at Kitt Peak where a 90Prime camera is installed at the prime focus of the telescope. The camera is composed of four 4096${\times}$4032 CCDs and provides a field of view of about 1 $deg^{2}$. The depths of the single-epoch images are about 23.4 mag and 22.9 mag for g and r, respectively. The typical exposure time is about 60 s, depending on the meteorological conditions. BASS, Dark Energy Camera Legacy Survey \citep{blum+16}, and MOSAIC z-band Legacy Survey \citep{silva+16} constitute the Dark Energy Spectroscopic Instrument (DESI) imaging surveys. BASS will survey about 5400 $deg^{2}$ in the northern Galactic cap and provide spectroscopic targets for DESI. Besides, BASS will also provide unique science opportunities for studying Galactic halo substructures, satellite dwarf galaxies around the Milky Way, galaxy clusters, high-redshift quasars, and so on \citep{zou+etal+2017+aj}. Table~\ref{tab0} summarizes the specifications of the Bok telescope and the CCD detector.

\begin{table}
\caption{\label{tab0}Specifications of the Bok telescope and CCD detector} \centering
\begin{tabular}{rr}
\hline
Telescope                         &2.3 m Bok\\
\hline
F-Ratio                           &f/2.98\\
Diameter of primary mirror        &230cm\\
Absolute pointing                 &<3${\arcsec}$ rms\\
CCD field of view            &1.08${\degr}$${\times}$1.03${\degr}$\\
Size of CCD array            &4096${\times}$4032\\
Size of pixel               &${15\mu}{\times}{15\mu}$\\
Approximate angular extent per pixel  &0${\farcs}$454/pixel\\
Typical exposure time          &60 s\\
\hline
\end{tabular}
\end{table}

\begin{figure}
   \centering
   \includegraphics[width=8cm, angle=0]{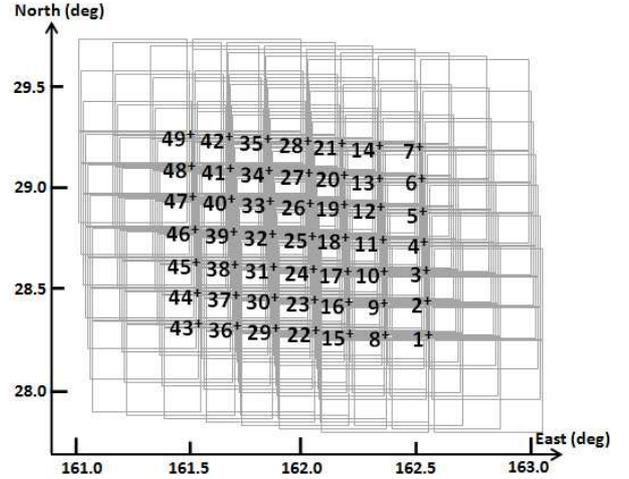}
   \caption{Dithered observational scheme for the CCD frames in 2016. The 49 images are organized in a 7${\times}$7 array. The unit is degree. The observational scheme for 2017 is similar to that of 2016.}
   \label{Fig1}
\end{figure}

However, the astrometry of BASS is unsatisfactory. The astrometric solutions of BASS are derived by cross-identifying objects in the frames with SDSS and 2MASS star catalogue \citep{zhou+16}, during the first data release of BASS, the median positional errors of stars from 9000 frames in right ascension and declination are about 0.10 arcsec and 0.11 arcsec, respectively \citep{zou+etal+2017+aj}. There are several factors that affect the accuracy of astrometric solution, such as the precision of the source positional measurements, the solution of the GD, the positional precision of the reference stars, and so on. Among these factors, the solution of the GD is an important one. In general, there are two ways to derive the GD. The direct way would be to observe an astrometric flat field where we have the prior knowledge of the positions of all the stars in some absolute and accurate system \citep{Anderson+2003}. Then the distortion would be obtained by means of the residuals. However, such standard calibration field is usually not available due to the lack of precision and star density. Although Gaia Data Release 1 (DR1) \citep{gaia+16b,gaia+16a} can provide position precision and star density, the lack of proper motion for most of stars can induce negligible errors. With the release of Gaia DR2 \citep{gaia+2018}, it can be expected that Gaia DR2 should achieve good results. Another way is to take advantage of the instrument itself to calibrate. Anderson et al. (\citeyear{Anderson+2006}) applied the self-calibration method to the wide-field ground-based images and achieved good precision. Peng et al. (\citeyear{Peng+etal+2012}) presented an alternative self-calibration method and successfully applied the method to the observations of some natural satellites \citep{Peng+etal+2015,Wang+etal+2017,Peng+etal+2017}. Therefore, self-calibration method is more practical and effective as the derivation of the GD is free from the errors of the star catalogue in positions and proper motions, but additional observations for the calibration field should be made. According to Peng et al. (\citeyear{Peng+etal+2012}), the observation of the calibration field should be made at different offset in a dithered pattern of either "+" or "$\#$". The offset between any two neighboring exposures should be appropriate and the size of the offset depends on the change of the GD pattern. The field of view of the Bok telescope is quite wide and thus much requirement should be met for the offset (such as more reference stars, smaller offsets and so on). However, during the exposures of the same sky field for BASS, the offset between two neighboring exposures are about 10 arcmin in right ascension or in declination, see Figure~\ref{Fig1}. Due to the insufficient overlapping area, the derived lookup table is inadequate and the positional precision of the star is affected. Then two third-order polynomials were tried beyond the numerical GD model. The results show that the positional precision of the stars is improved greatly after the third-order polynomial GD correction. Here, the averaged GD in a small box (see section 3.1) is called a numerical GD and the GD derived from a polynomial is called an analytical GD. Moreover, a precise pixel positional measurement is fundamental. During the pixel positional measurement, two dimensional Gauss fit was used to get the precise positions of the stars \citep{liz+09} as we have done in our previous work \citep{Peng+etal+2015,Wang+etal+2015,Wang+etal+2017}. Besides, in order to correct all the smaller-scale systematic positional errors, an additional lookup table correction was found useful.

The contents of this paper are organized as follows. In Section 2, the observations are described. Section 3 presents the solution of the GD models. In Section 4, we make analysis of the relative positions of the chips. Finally, in Section 5, conclusions are drawn.

\section{Observations with different filters}
\label{sect:Obs}
\begin{figure}
   \centering
   \includegraphics[width=6cm, angle=0]{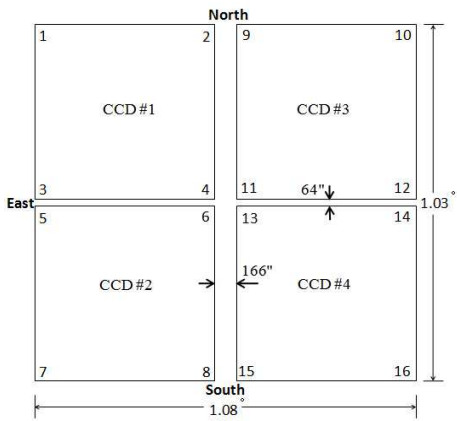}
   \caption{Layout of the CCD array. There are four CCDs: CCD$\#$1, CCD$\#$2, CCD$\#$3, CCD$\#$4 and four identifiers are listed at the corners of each CCD. An exposure includes four CCD frames. (see Figure 1 in \citealt{zou+etal+2017+aj})}
   \label{Fig2}
\end{figure}

Before observation in each night, the x-axis of each CCD chip is nearly parallel to the real equator by a star-trailing operation. Then two sets of the observations were taken in 2016 and 2017, see Table~\ref{tab1}. All the observations were made with the 2.3 m Bok telescope at Kitt Peak. The site (i.e. IAU code 691) is at longitude E248${\degr}$ 24${\arcmin}$ 3.6${\arcsec}$, latitude N31${\degr}$ 57${\arcmin}$ 28.7${\arcsec}$ and height 2079.8 m above sea level. The exposure time for each CCD frame is 60 s and the pixel scale is about 0.454${\arcsec}$. The observation in 2016 were made at high galactic latitude and in each frame a small number of reference stars ($\sim$~300 stars) are available. The center of the field of observation is at E162.05${\degr}$ in right ascension, N28.65${\degr}$ in declination. The observation in 2017 were made at low galactic latitude and a plenty of reference stars ($\sim$~3200 stars) are available in each frame. The center of the field is at E147.35${\degr}$ in right ascension, N1${\degr}$ in declination. An exposure includes four frames and a total of 4${\times}$87 frames of CCD images have been obtained in a dithered pattern (see Figure~\ref{Fig1}). There are inter-chip gap in the center of the array, see Figure~\ref{Fig2} (\citealt{zou+etal+2017+aj}). The Bok telescope has an equatorial mount and other information about the telescope and the observations could be found from the FITS header. A sample FITS header could be seen in Appendix.

\begin{table}
\centering
\begin{minipage}[]{90mm}
\caption[]{Observations for the calibration field. Column 2 lists the number of CCD frames and column 3 lists the filter.\label{tab1}}\end{minipage}
\setlength{\tabcolsep}{1pt}
\small
 \begin{tabular}{ccccccc}
 \\
  \hline\noalign{\smallskip}
Obs Date&Calibration fields No.&filter\\
  \hline\noalign{\smallskip}
2016-01-17 & 4${\times}$49&r\\
2017-03-05 & 4${\times}$38&g\\
  \noalign{\smallskip}\hline
\end{tabular}
\end{table}

\section{Solve for the GD models progressively}
The reduction procedures were carried out according to \citet{Peng+etal+2012}. First, a numerical GD model was derived. Then an analytical GD model was tried as the residuals of the stars are still large after the numerical GD correction. After the analytical GD correction, the positional precision of stars has been improved greatly. Finally, an additional lookup table is set up as the residual GD does not converge for the preset threshold. More explanations are described as follows.
\subsection{The derivation of the numerical GD model}
\label{sect:Red}
\begin{figure}
\centering
\includegraphics[width=8.5cm, angle=0]{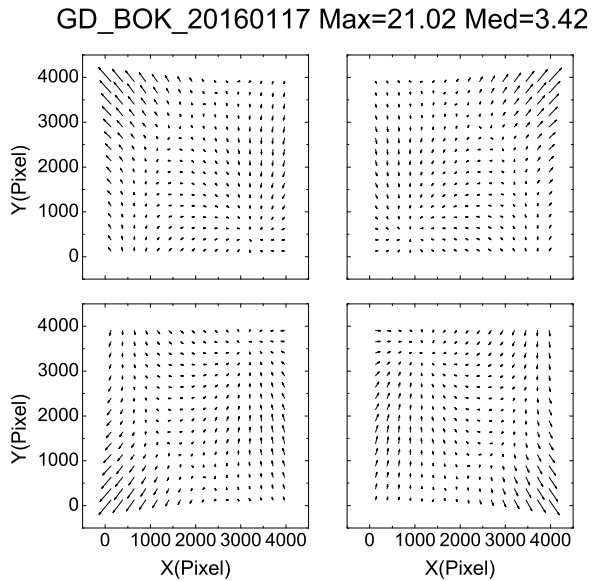}
\caption{GD for the Bok 2.3 m telescope. The GD was derived from the observations in 2016 and r- filter was used. The maximum GD (Max) and the median GD (Med) are listed in units of pixels and a factor of 20 is used to exaggerate the magnitude of the GD vectors.}\label{FigGD16}
\end{figure}

\begin{figure}
\centering
\includegraphics[width=8.5cm, angle=0]{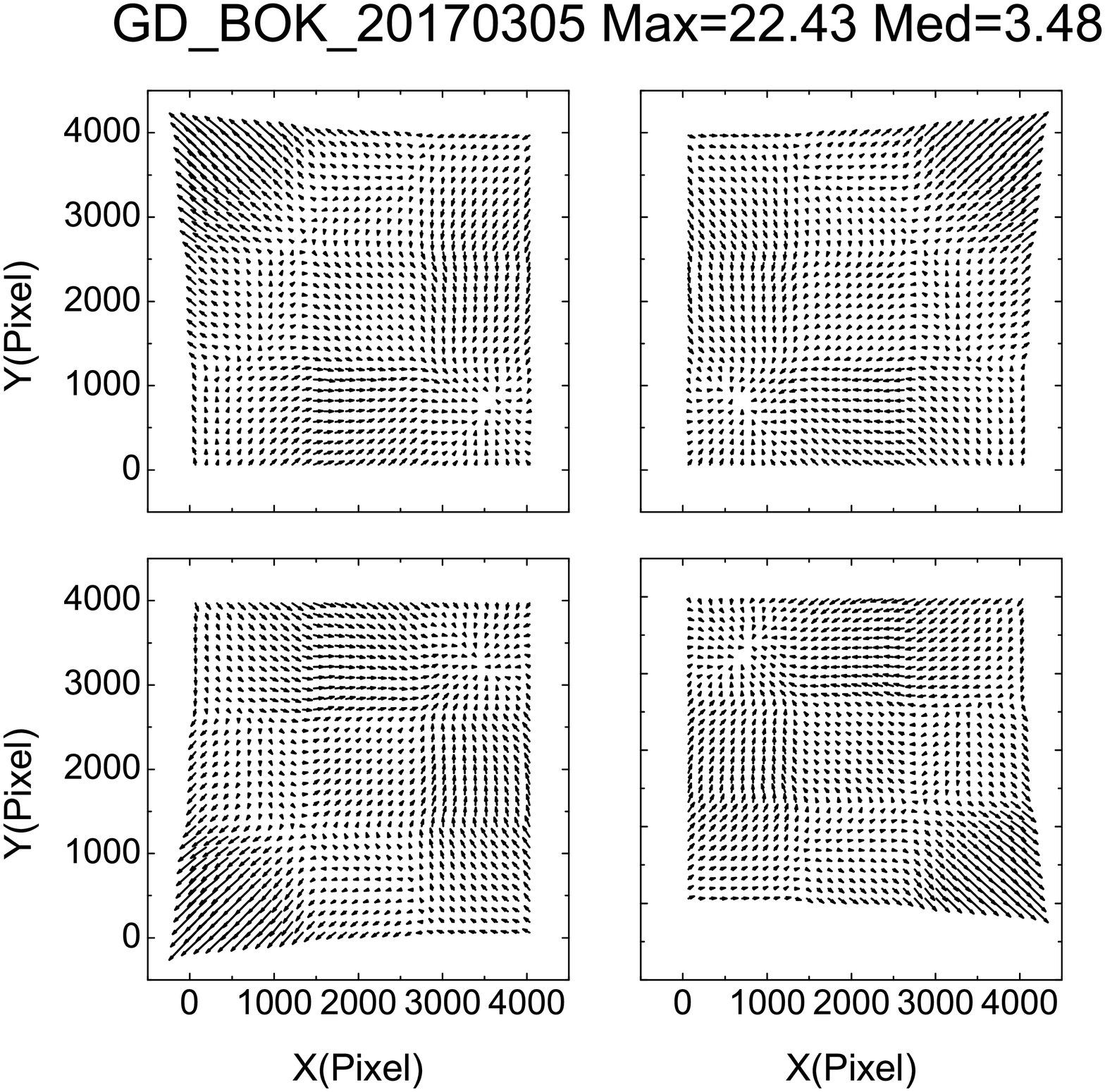}
\caption{GD for the Bok 2.3 m telescope. The GD was derived from the observations in 2017 and g- filter was used. The maximum GD (Max) and the median GD (Med) are listed in units of pixels and a factor of 20 is used to exaggerate the magnitude of the GD vectors.}\label{FigGD17}
\end{figure}

\begin{figure}
\centering
\includegraphics[width=8.5cm, angle=0]{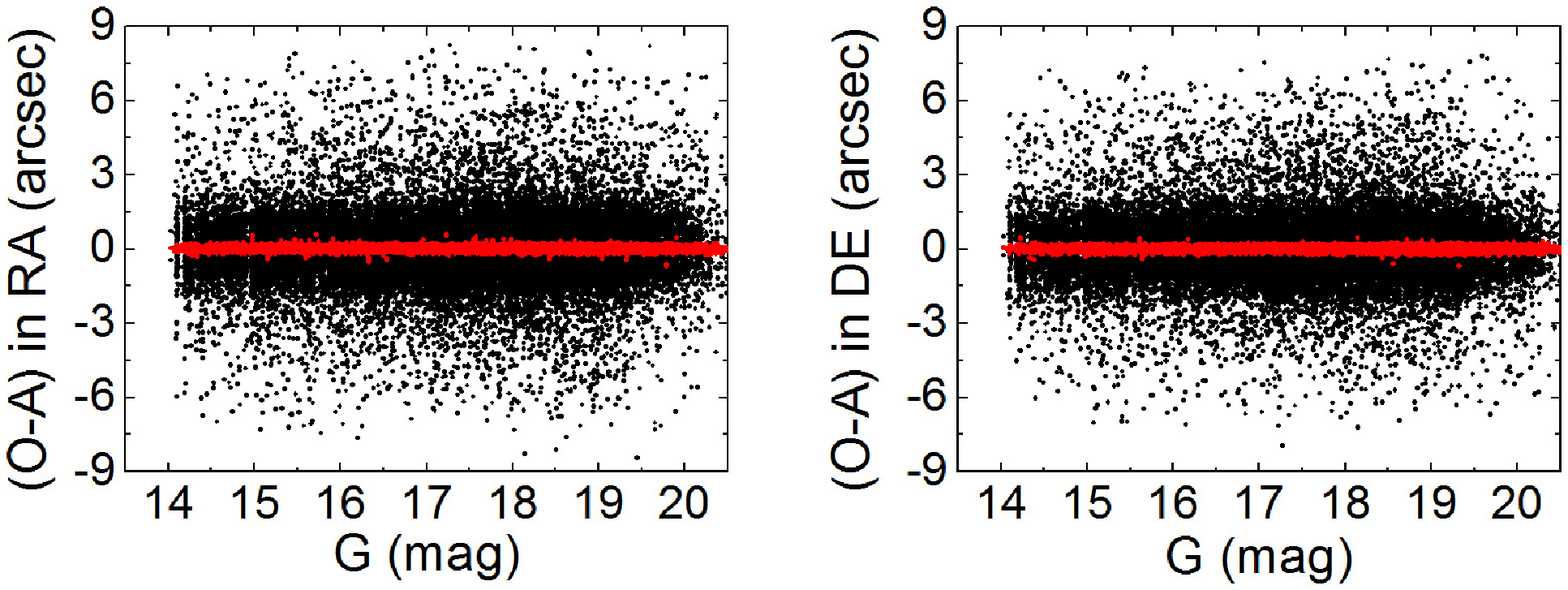}
\caption{(O-A) residuals of the stars in 2016. The dark points represent the residuals before the numerical GD correction where four parameters transformation is used and the red ones represent the residuals after the numerical GD correction.}\label{Fig5}
\end{figure}

\begin{figure}
\centering
\includegraphics[width=8.5cm, angle=0]{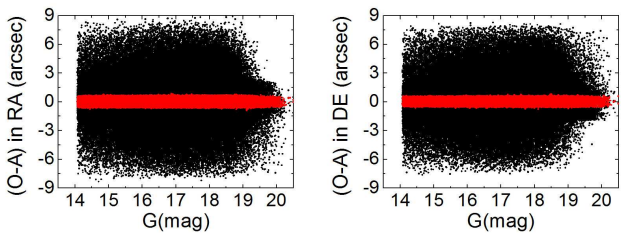}
\caption{(O-A) residuals of the stars in 2017. The dark points represent the residuals before the numerical GD correction where four parameters transformation is used and the red ones represent the residuals after the numerical GD correction.}\label{Fig6}
\end{figure}

The derivation of the numerical GD model mainly involves the following steps: (1) Firstly, some information about the date of the observation, exposure time, and so on are extracted automatically from the FITS header by our own software (\citealt{Peng+etal+2017,Wang+etal+2017}). Then after some preprocessing (de-bias, flat correction), the pixel positions of the stars in each CCD frame are measured with two dimensional Gauss fit and this is different from the past when SExtractor (\citealt{Bertin+1996}) was used for the pixel positional measurement. As the solution of the GD can be affected by the precision of those stars too bright (brighter than 14 mag and saturated), these stars will not be considered. For each CCD frame in 2016, about 100 stars are rejected and about 200 stars are used. For each CCD frame in 2017, about 200 stars are rejected and about 3000 stars are used. Then the stars in each frame are matched with the stars in Gaia DR2 star catalogue by using some fast matching algorithm (\citealt{ren+2010}). The pixel positional measurement and the match process are performed automatically by our own software. (2) For each chip we adopted the center of the CCD array as the tangential point on the tangential plane of the image. Then the standard coordinates of the stars are obtained by the central projection (\citealt{green1985}). Next, the positional residuals (observed minus averaged; O-A) of the stars can be obtained after four parameters transformation. As an accurate GD model requires high precision of the stars positions, then during the computations of the standard coordinate, the topocentric apparent position and atmospheric refraction for each matched star in each CCD frame are also taken into account. (3) Due to the plenty of reference stars available in 2017, each CCD frame is divided into 1024 equal-area boxes and the size of each box is 128${\times}$126 pixels. Each CCD frame in 2016 is divided into 256 boxes and the size is 256${\times}$252 pixels for each box. Afterward the averaged GD in each box was obtained by cancelling out the catalogue errors and compressing the measured errors. By ten to twelve iterations the numerical GD converged and was derived out. In fact the derived numerical GD model is a lookup table. Figure~\ref{FigGD16} and Figure~\ref{FigGD17} show the numerical GD patterns for the Bok telescope in 2016 and 2017, respectively. We can see that the telescope suffers a serious GD, especially in the corners. The maximum GD is about 22 pixels ($\sim$~10.18 arcsec). Figure~\ref{Fig5} and Figure~\ref{Fig6} show the positional residuals of the stars with respect to the magnitude in 2016 and 2017, respectively. We can see the positional residuals are suppressed largely after the numerical GD correction.

\subsection{The derivation of the analytical GD model}
\begin{figure}
\centering
\includegraphics[width=8.5cm, angle=0]{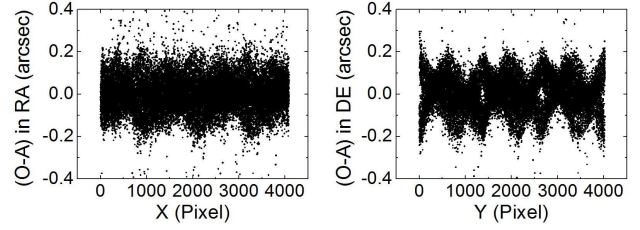}
\caption{Positional residuals of the stars with respect to the pixel positions in 2016.}\label{Fig5_1}
\end{figure}

\begin{figure}
\centering
\includegraphics[width=8.5cm, angle=0]{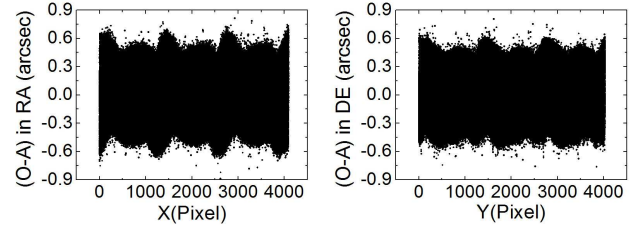}
\caption{Positional residuals of the stars with respect to the pixel positions in 2017.}\label{Fig5_2}
\end{figure}

By using the numerical GD correction, the precision of the stars is improved. However, the positional residuals of the stars are still large. Specifically, we found the residuals did not distribute randomly with respect to the stars' pixel positions (see Figure~\ref{Fig5_1} and Figure~\ref{Fig5_2}). Besides, the GD pattern in 2017 seems to be divided into several blocks (see Figure~\ref{FigGD17}). Then the analytical GD model was tried beyond the numerical GD model.

As we know, the previous GD pattern is a numerical model and each derived GD is an average in a box with a size of 128${\times}$126 or 256${\times}$252 pixels. Then the averaged GD in the small box cannot represent the GD pattern finely due to the serious GD pattern. In order to depict the GD pattern in more details, two third-order polynomials were used to fit the previous numerical GD. Higher-order polynomials were also tried, but the results show that the positional residuals of the stars are almost the same as that of third-order polynomial fit. So the third-order polynomial fit was chosen. The positional residuals of the stars for higher-order polynomial fit will be given later (see Table 5). Here, the detailed third-order polynomials are shown by the following equations:
\begin{equation}\label{poly}
\begin{aligned}
{\Delta X} &= a_{1}x^3 + a_{2}x^2y + a_{3}xy^2 + a_{4}y^3+a_{5}x^2+a_{6}xy+a_{7}y^2\\
& +a_{8}x+a_{9}y+a_{10}\\
{\Delta Y} &= b_{1}x^3 + b_{2}x^2y + b_{3}xy^2 + b_{4}y^3+b_{5}x^2+b_{6}xy+b_{7}y^2\\
& +b_{8}x+b_{9}y+b_{10}
\end{aligned}
\end{equation}
where ($\mit \Delta X$, $\mit \Delta Y$) is the solved GD for the center of each box, ($\mit x$, $\mit y$) is its normalized pixel coordinate between -1 and 1, $\mit a$$_{i}$ and $\mit b$$_{i}$ ($\mit i$=1 to 10) are the coefficients of the polynomials. We use the normalized pixel positions:
\begin{center}
$x=(x_{obs}-2048)/2048$\\
$y=(y_{obs}-2016)/2016$
\end{center}
This normalization makes it easier to recognize the size of the contribution of each term to the solution.

After fitting, the coefficients of the polynomial were obtained and the analytical GD model was derived. Next an iterative procedure is necessary.

\begin{figure}
\centering
\includegraphics[width=8.5cm, angle=0]{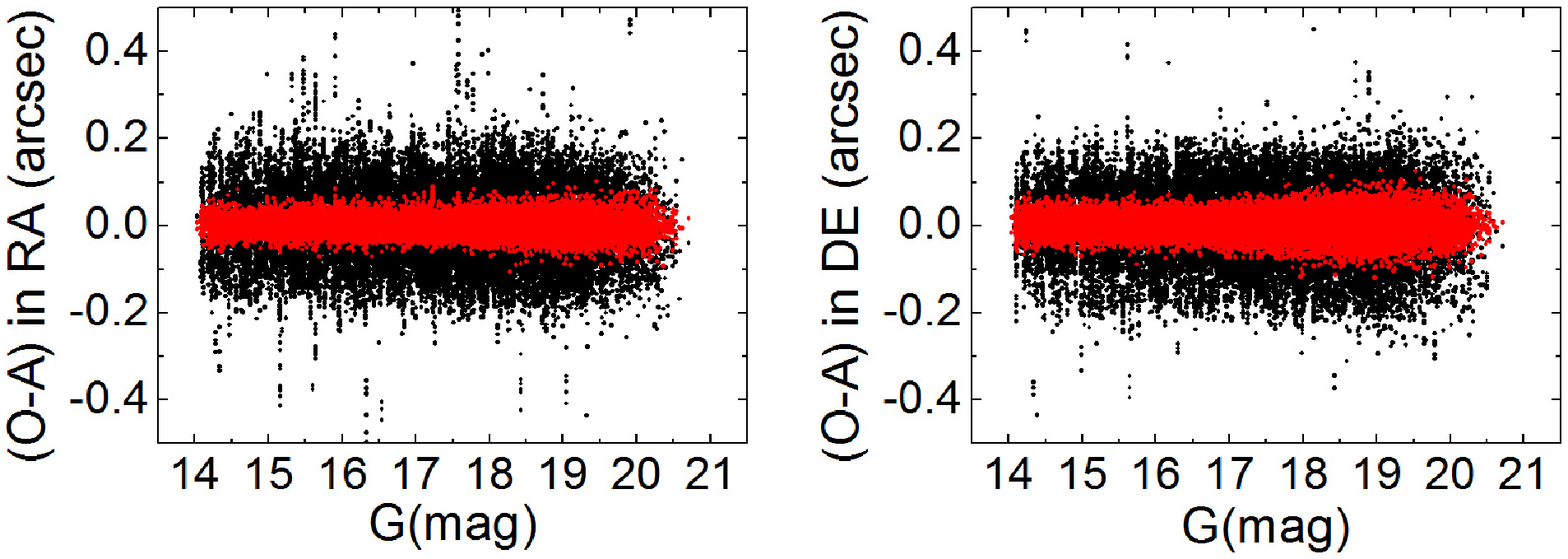}
\caption{Positional residuals of the stars with respect to the magnitude in 2016. The black points represent the residuals after the numerical GD correction and the red ones after the analytical GD correction.}\label{Fig6_1}
\end{figure}

\begin{figure}
\centering
\includegraphics[width=8.5cm, angle=0]{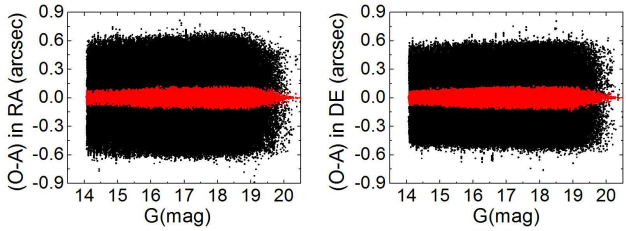}
\caption{Positional residuals of the stars with respect to the magnitude in 2017. The black points represent the residuals after the numerical GD correction and the red ones after the analytical GD correction.}\label{Fig6_2}
\end{figure}

Table~\ref{tab2} lists the solved coefficients of the third-order polynomials in 2016 and 2017, respectively and Figure~\ref{Fig6_1} and Figure~\ref{Fig6_2} show the positional residuals of the stars with respect to the magnitude. From Figure~\ref{Fig6_1} and Figure~\ref{Fig6_2} we can see that the positional precision of the stars is improved greatly. The final GD correction reaches a positional precision level of $\sim$~0.05 pixel ($\sim$~22 mas) in each coordinate and most of the GD was removed. Then the definition of precision of the GD correction can be given, that is, the precision of the GD correction can be denoted by the star positional residual after the GD correction.

\begin{table*}
\begin{minipage}[]{180mm}
\caption[]{The coefficient of the analytical model for the observations in 2016 and 2017. Column 1 lists the designation of each CCD chip and column 2 lists the date. Column 3 lists the direction ($\emph{a}$ or $\emph{b}$). Column 4 to 13 list the coefficients of the polynomials.\label{tab2}}
\end{minipage}
\centering
\setlength{\tabcolsep}{1pt}
\small
 \begin{tabular}{cccrrrrrrrrrr}
  \hline\noalign{\smallskip}
CCD&date&&1$\quad$$\quad$&2$\quad$$\quad$&3$\quad$$\quad$&4$\quad$$\quad$&5$\quad$$\quad$&6$\quad$$\quad$&7$\quad$$\quad$&8$\quad$$\quad$&9$\quad$$\quad$&10$\quad$\\
  \hline\noalign{\smallskip}
$\#$1&2016$\quad$&$\emph{a}$$\quad$& 1.897$\quad$& 0.090$\quad$& 1.857$\quad$& 0.020$\quad$&-6.256$\quad$&-3.953$\quad$&-2.039$\quad$&-0.658$\quad$& 2.335$\quad$& 2.754\\
     &2016$\quad$&$\emph{b}$$\quad$& 0.024$\quad$& 1.875$\quad$& 0.068$\quad$& 2.065$\quad$&-1.962$\quad$&-4.186$\quad$&-5.744$\quad$& 6.392$\quad$&-1.371$\quad$& 2.559\\
$\#$2&2016$\quad$&$\emph{a}$$\quad$& 1.880$\quad$&-0.068$\quad$& 1.856$\quad$&-0.021$\quad$&-6.264$\quad$& 3.694$\quad$&-2.041$\quad$& 0.038$\quad$&-2.298$\quad$& 2.757\\
     &2016$\quad$&$\emph{b}$$\quad$&-0.068$\quad$& 1.848$\quad$&-0.057$\quad$& 1.909$\quad$& 1.820$\quad$&-4.195$\quad$& 5.373$\quad$&-5.936$\quad$&-1.052$\quad$&-2.388\\
$\#$3&2016$\quad$&$\emph{a}$$\quad$& 1.889$\quad$&-0.090$\quad$& 1.855$\quad$&-0.037$\quad$& 5.871$\quad$&-3.973$\quad$& 1.913$\quad$&-1.820$\quad$&-1.426$\quad$&-2.585\\
     &2016$\quad$&$\emph{b}$$\quad$&-0.027$\quad$& 1.874$\quad$&-0.085$\quad$& 1.904$\quad$&-1.978$\quad$& 3.896$\quad$&-5.799$\quad$&-6.813$\quad$&-1.791$\quad$& 2.582\\
$\#$4&2016$\quad$&$\emph{a}$$\quad$& 1.886$\quad$& 0.087$\quad$& 1.865$\quad$& 0.035$\quad$& 5.906$\quad$& 3.687$\quad$& 1.937$\quad$&-0.153$\quad$& 2.474$\quad$&-2.604\\
     &2016$\quad$&$\emph{b}$$\quad$& 0.036$\quad$& 1.878$\quad$& 0.082$\quad$& 1.910$\quad$& 1.831$\quad$& 3.932$\quad$& 5.354$\quad$& 5.114$\quad$&-0.825$\quad$&-2.385\\
$\#$1&2017$\quad$&$\emph{a}$$\quad$& 1.942$\quad$& 0.084$\quad$& 1.857$\quad$& 0.015$\quad$&-6.333$\quad$&-4.049$\quad$&-2.044$\quad$&-1.198$\quad$& 4.493$\quad$& 2.790\\
     &2017$\quad$&$\emph{b}$$\quad$& 0.023$\quad$& 1.885$\quad$& 0.074$\quad$& 1.889$\quad$&-2.020$\quad$&-4.200$\quad$&-5.877$\quad$& 4.389$\quad$&-1.613$\quad$& 2.630\\
$\#$2&2017$\quad$&$\emph{a}$$\quad$& 1.926$\quad$&-0.030$\quad$& 1.867$\quad$&-0.005$\quad$&-6.308$\quad$& 3.668$\quad$&-2.025$\quad$&-0.383$\quad$&-4.012$\quad$& 2.775\\
     &2017$\quad$&$\emph{b}$$\quad$&-0.002$\quad$& 1.881$\quad$&-0.027$\quad$& 1.873$\quad$& 1.813$\quad$&-4.150$\quad$& 5.364$\quad$&-4.091$\quad$&-1.589$\quad$&-2.390\\
$\#$3&2017$\quad$&$\emph{a}$$\quad$& 1.922$\quad$&-0.032$\quad$& 1.829$\quad$&-0.019$\quad$& 5.987$\quad$&-4.069$\quad$& 1.929$\quad$&-1.559$\quad$&-4.398$\quad$&-2.636\\
     &2017$\quad$&$\emph{b}$$\quad$&-0.009$\quad$& 1.861$\quad$&-0.031$\quad$& 1.830$\quad$&-2.051$\quad$& 3.919$\quad$&-5.922$\quad$&-4.093$\quad$&-1.329$\quad$& 2.655\\
$\#$4&2017$\quad$&$\emph{a}$$\quad$& 1.911$\quad$& 0.052$\quad$& 1.878$\quad$& 0.007$\quad$& 6.010$\quad$& 3.643$\quad$& 1.941$\quad$&-0.852$\quad$& 3.845$\quad$&-2.648\\
     &2017$\quad$&$\emph{b}$$\quad$& 0.013$\quad$& 1.898$\quad$& 0.049$\quad$& 1.832$\quad$& 1.810$\quad$& 3.913$\quad$& 5.316$\quad$& 3.663$\quad$&-1.612$\quad$&-2.373\\
  \noalign{\smallskip}\hline
\end{tabular}
\end{table*}

\subsection{Setting up an additional lookup table}
\begin{figure}
\centering
\includegraphics[width=8.5cm, angle=0]{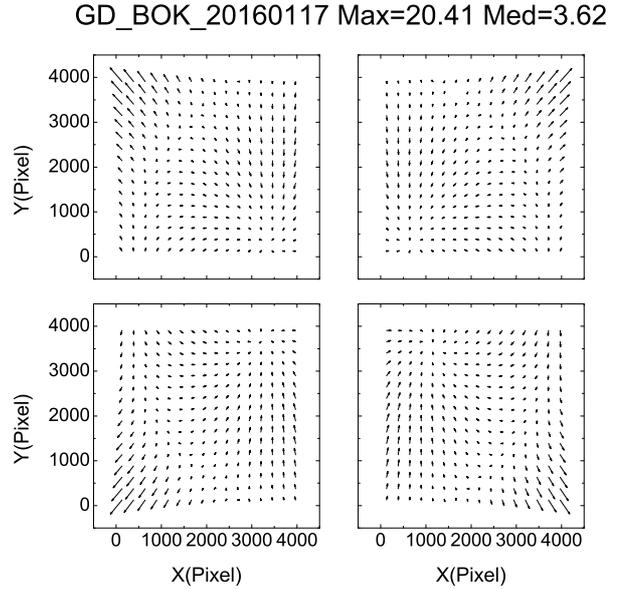}
\caption{GD for the Bok 2.3 m telescope in 2016. The GD was derived from the analytical GD model and the additional lookup table in 2016. The maximum GD (Max) and the median GD (Med) are listed in units of pixels and a factor of 20 is used to exaggerate the magnitude of the GD vectors.} \label{Fig6_3}
\end{figure}

\begin{figure}
\centering
\includegraphics[width=8.5cm, angle=0]{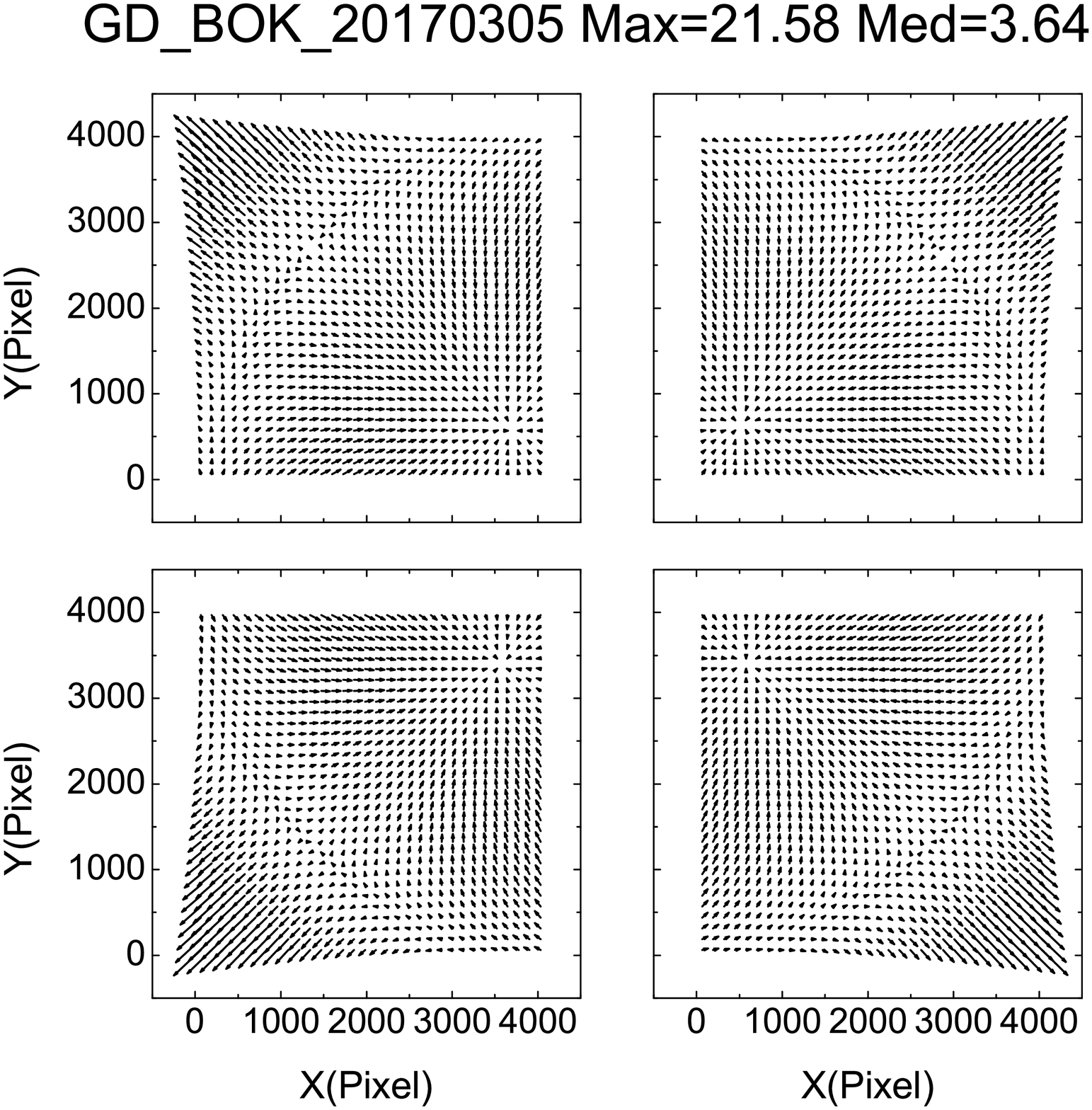}
\caption{GD for the Bok 2.3 m telescope in 2017. The GD was derived from the analytical GD model and the additional lookup table in 2017. The maximum GD (Max) and the median GD (Med) are listed in units of pixels and a factor of 20 is used to exaggerate the magnitude of the GD vectors.}\label{Fig6_4}
\end{figure}

\begin{figure}
\centering
\includegraphics[width=8.5cm, angle=0]{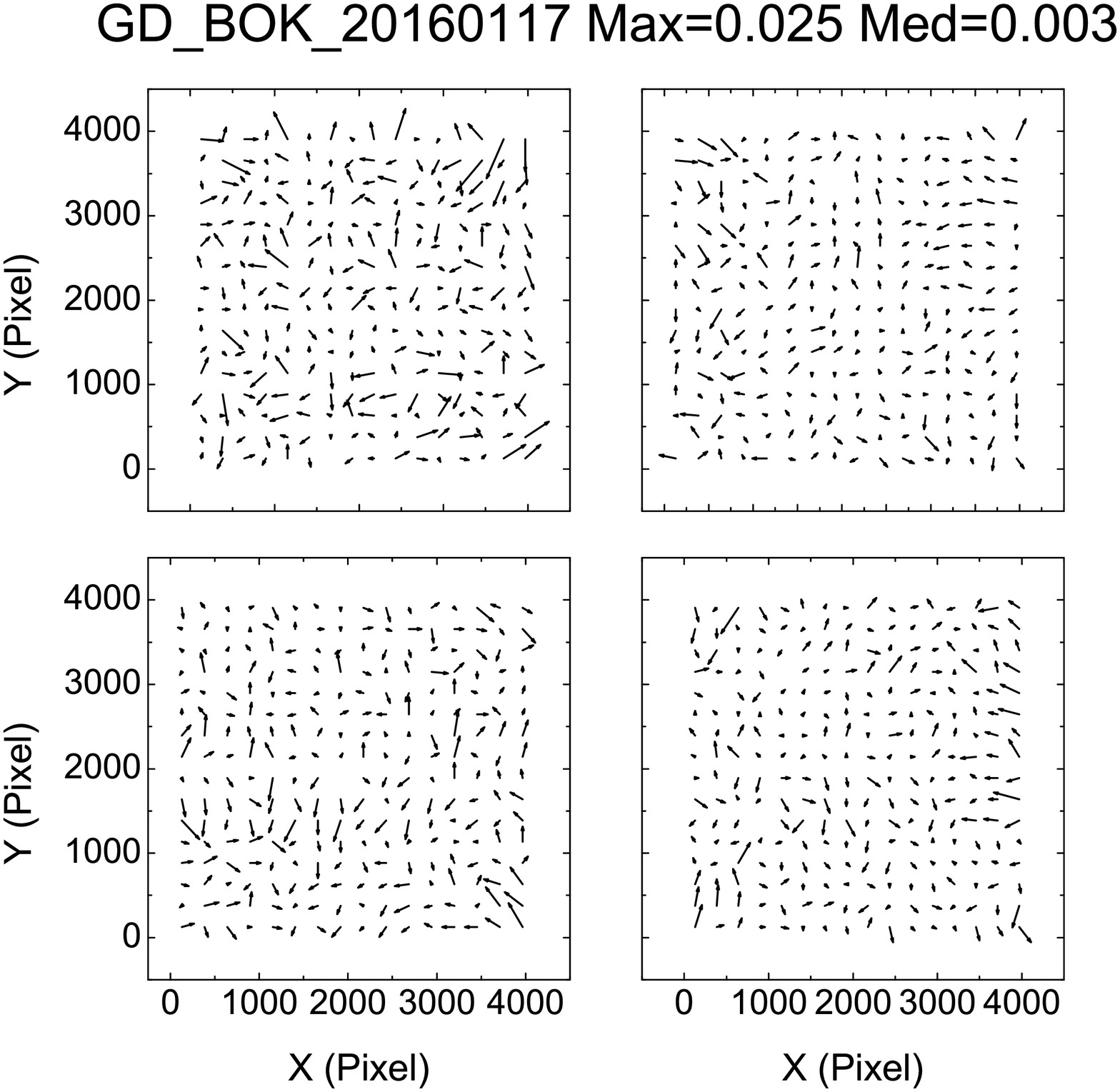}
\caption{Residuals for the Bok 2.3 m telescope in 2016. The maximum GD (Max) and the median GD (Med) are listed in units of pixels and a factor of 20000 is used to exaggerate the magnitude of the GD vectors.} \label{Fig11_1}
\end{figure}

\begin{figure}
\centering
\includegraphics[width=8.5cm, angle=0]{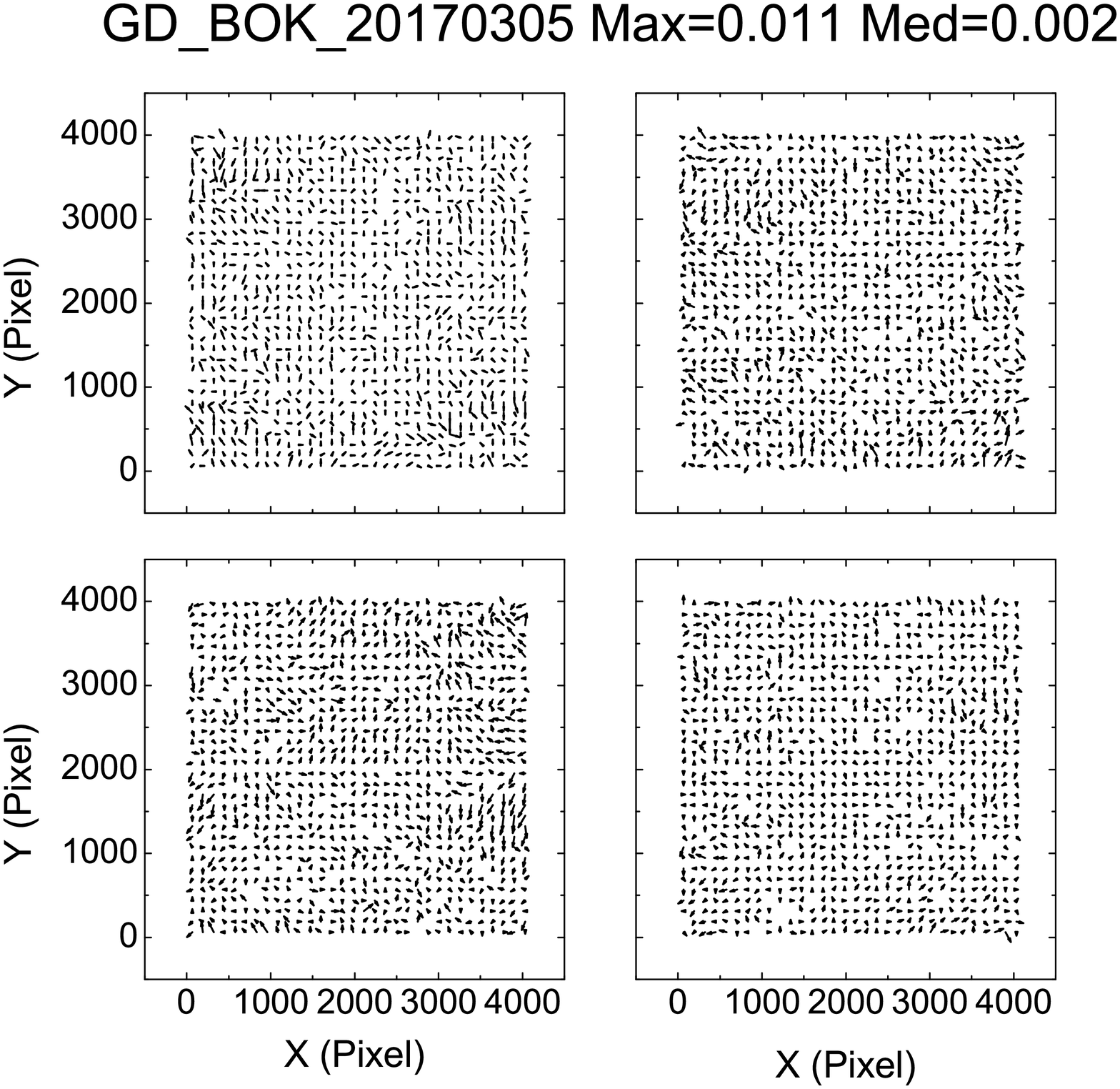}
\caption{Residuals for the Bok 2.3 m telescope in 2017. The maximum GD (Max) and the median GD (Med) are listed in units of pixels and a factor of 20000 is used to exaggerate the magnitude of the GD vectors.}\label{Fig11_2}
\end{figure}

Although the precision for a star's reduced position is greatly improved after the analytical GD correction, during the iterative solution of the analytical GD model, the numerical GD does not converge for the preset threshold, for example, 0.01 pixel. Therefore an additional lookup table was set up and its corresponding lookup table correction was followed after the analytical GD correction. After the lookup table correction, the numerical GD converged quickly during the iterative solution. Figure~\ref{Fig6_3} and Figure~\ref{Fig6_4} show the analytical GD patterns for the Bok telescope in 2016 and 2017, respectively. From Figure~\ref{Fig6_4} we can see that the blocks found in Figure~\ref{FigGD17} have disappeared. In order to show the result of the GD correction, the final residual patterns after the analytical GD correction and the lookup table correction were also shown in Figure~\ref{Fig11_1} and Figure~\ref{Fig11_2} where the magnitude of the residual is magnified by a factor of 20000.

\begin{figure}
\centering
\includegraphics[width=8.5cm, angle=0]{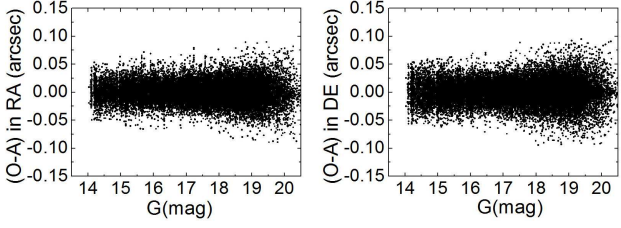}
\caption{Positional residuals of the stars with respect to the magnitude in 2016 in right ascension and declination, respectively. The black points represent the residuals of the stars after the analytical GD correction and the additional lookup table correction.}\label{Fig7_1}
\end{figure}

\begin{figure}
\centering
\includegraphics[width=8.5cm, angle=0]{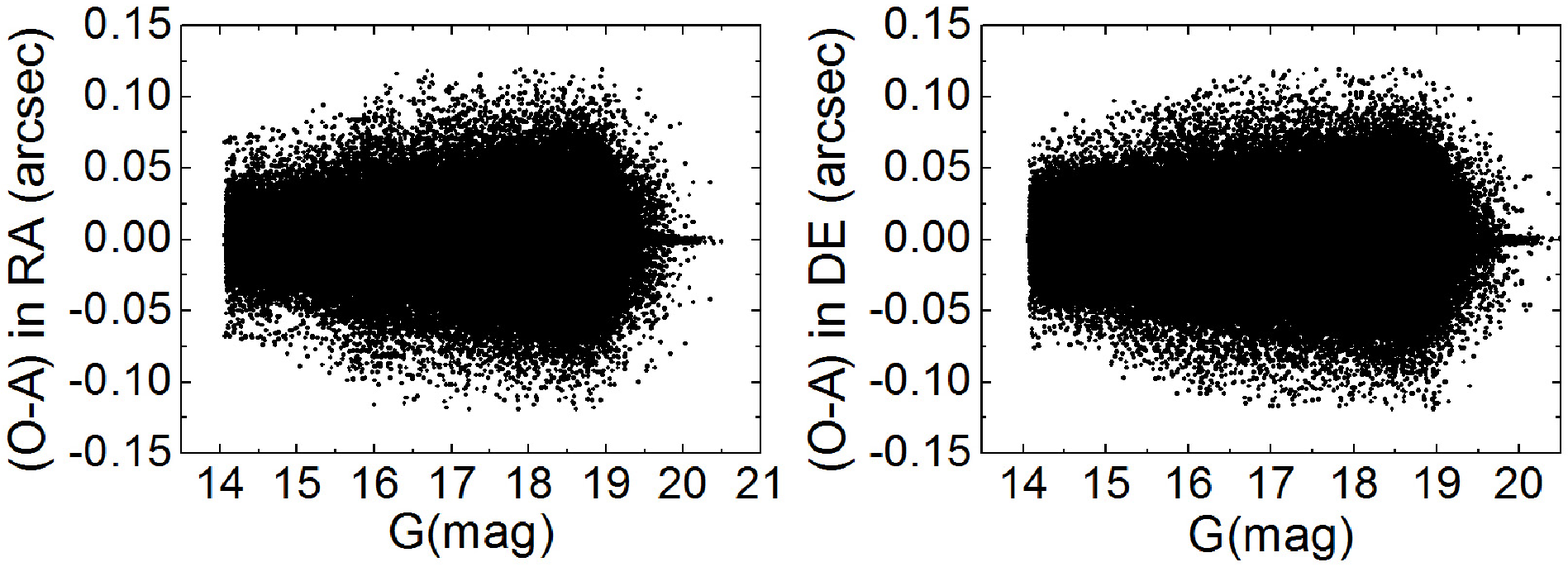}
\caption{Positional residuals of the stars with respect to the magnitude in 2017 in right ascension and declination, respectively. The black points represent the residuals of the stars after the analytical GD correction and the additional lookup table correction.}\label{Fig7_2}
\end{figure}

\begin{figure}
\centering
\includegraphics[width=8.5cm, angle=0]{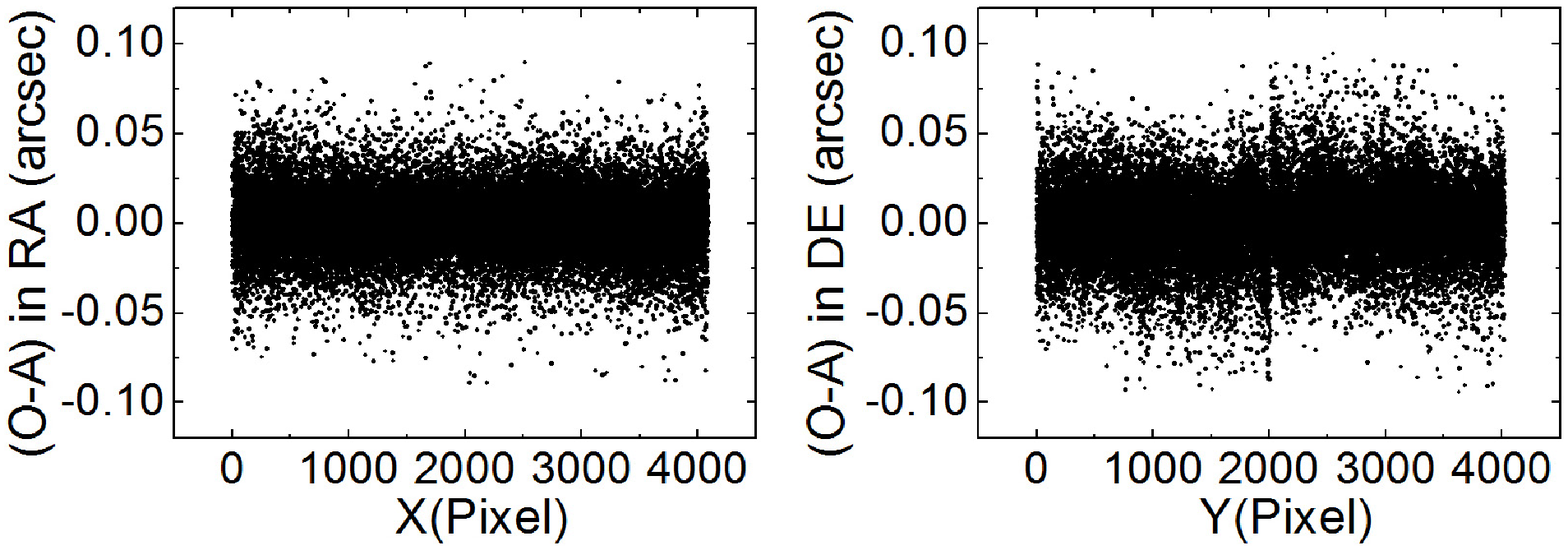}
\caption{Positional residuals of the stars with respect to the pixel positions in 2016. The dark points represent the residuals after the analytical GD correction and the additional lookup table correction.}\label{Fig8_1}
\end{figure}

\begin{figure}
\centering
\includegraphics[width=8.5cm, angle=0]{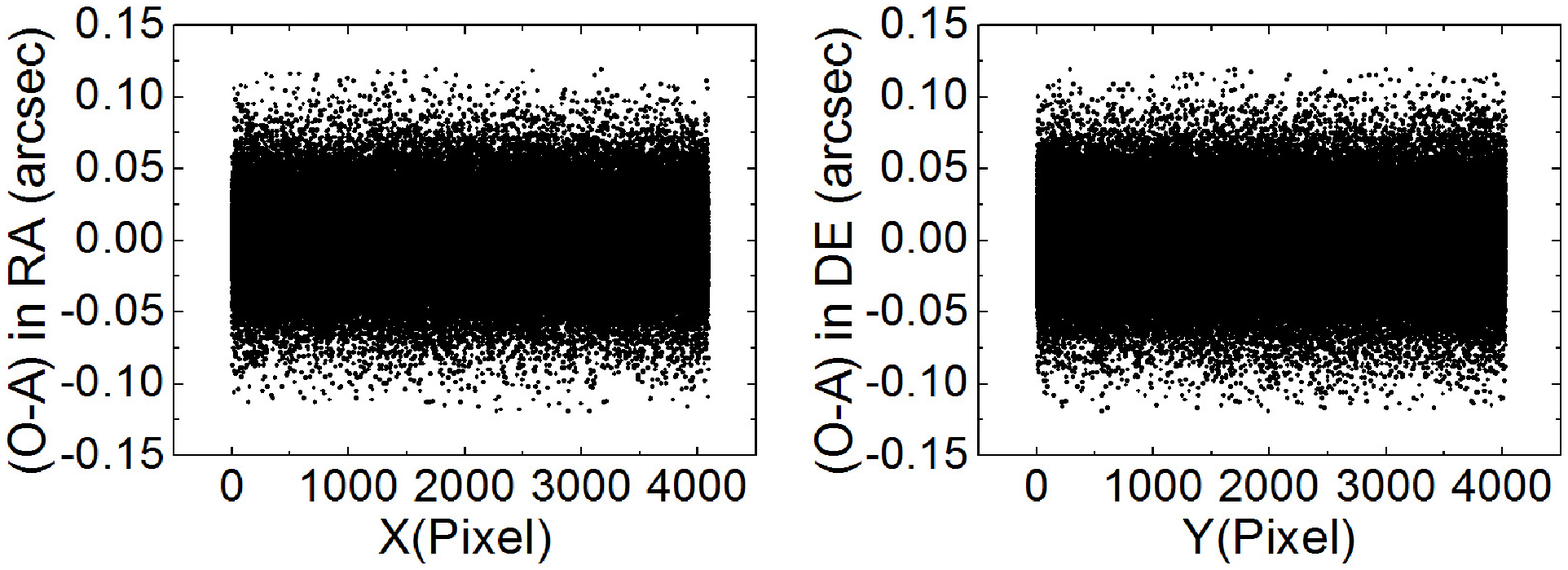}
\caption{Positional residuals of the stars with respect to the pixel positions in 2017. The dark points represent the residuals after the analytical GD correction and the additional lookup table correction.}\label{Fig8_2}
\end{figure}

\begin{figure}
\centering
\centering
\includegraphics[width=8.5cm, angle=0]{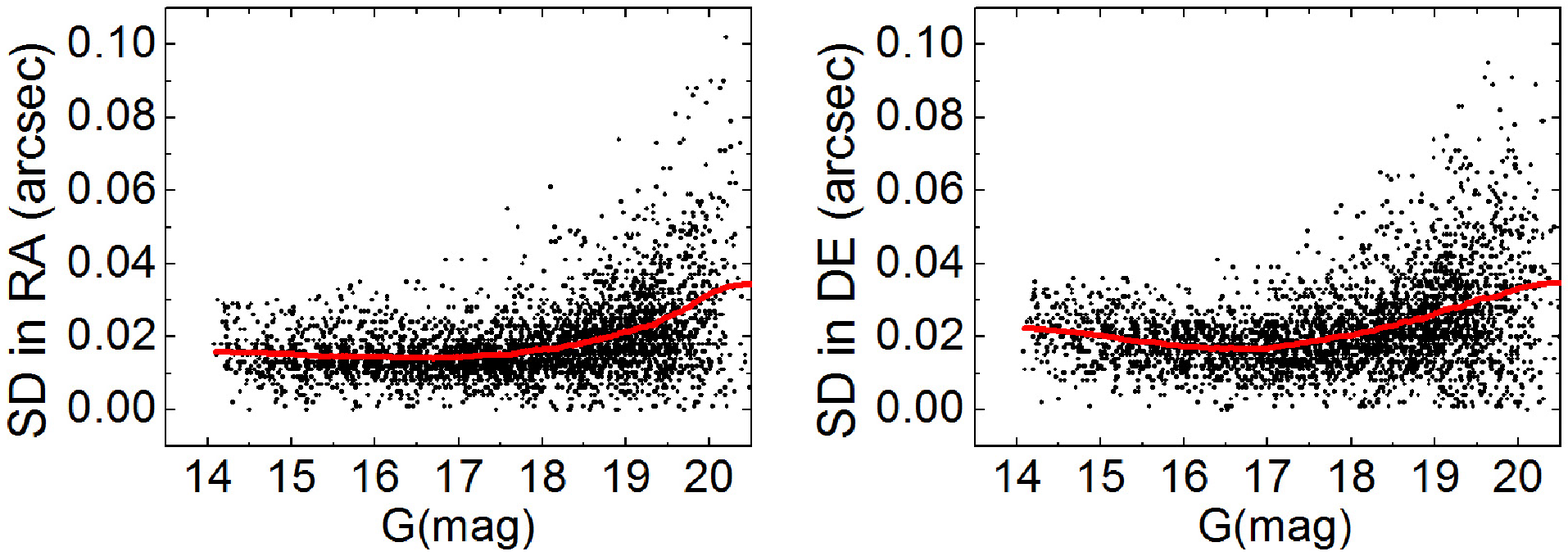}
\caption{Standard deviation of the positional residual for each star after the analytical GD correction and the additional lookup table correction with respect to the magnitude in 2016. A Savitzky-Golay smooth line is drawn for every 500 points.}\label{Fig9_1}
\end{figure}

\begin{figure}
\centering
\centering
\includegraphics[width=8.5cm, angle=0]{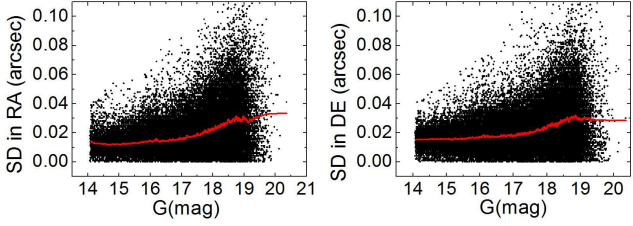}
\caption{Standard deviation of the positional residual for each star after the analytical GD correction and the additional lookup table correction with respect to the magnitude in 2017. A Savitzky-Golay smooth line is drawn for every 500 points.}\label{Fig9_2}
\end{figure}

Besides, Figure~\ref{Fig7_1} and Figure~\ref{Fig7_2} show the positional residuals of the stars after the analytical GD correction and the additional lookup table correction. Table~\ref{tab3} shows the mean and standard deviation of the stars's positional residuals by using the different GD models. From Figure~\ref{Fig7_1}, Figure~\ref{Fig7_2} and Table~\ref{tab3} we can see that the positional precision of the stars is further improved. Furthermore, the positional residuals of the stars with respect to the pixel positions are also shown in Figure~\ref{Fig8_1} and Figure~\ref{Fig8_2}. From these two figures, we can see that the positional residual distributes almost randomly after the analytical GD correction and the additional lookup table correction. However, a small hopping is seen in the right panel of Figure~\ref{Fig8_1}, and further study needs to be made in future. Table~\ref{tab4} shows the mean and standard deviation of the stars's positional residuals by using different orders polynomial fit. Figure~\ref{Fig9_1} shows the standard deviation of each star's positional residual in 2016 in right ascension and declination, respectively and Figure~\ref{Fig9_2} shows the standard deviation in 2017. In Figure~\ref{Fig9_1} and Figure~\ref{Fig9_2}, the stars that appeared more than twice are included and a smooth line is also drawn for every 500 points. In short, from these figures we can see that after the GD correction, the positional measurement precision for the bright stars is better than 20 mas. The positional measurement precision in 2017 is even better for most stars.

\begin{table}
\centering
\begin{minipage}[]{90mm}
\caption[]{Statistics of the positional residuals for all stars by using different GD models for each data set. Column 1 is the date. The second column shows the GD model. The following columns list the mean positional residual and its standard deviation (SD) in right ascension and declination, respectively. All units are in arcseconds.\label{tab3}}
\end{minipage}
\setlength{\tabcolsep}{1pt}
\small
 \begin{tabular}{ccccccccccccc}
 \\
  \hline\noalign{\smallskip}
  Date&  GD model          &$<$O-A$>$& SD & $<$O-A$>$& SD \\
      &                    &\multicolumn{2}{c}{RA}&\multicolumn{2}{c}{Dec.} \\
  \hline\noalign{\smallskip}
2016-01-17&Analytical&0.000&0.019&0.000&0.023\\
          &Analytical+lookup table&0.000&0.018&0.000&0.021\\
2017-03-05&Analytical&0.000&0.019&0.000&0.021\\
          &Analytical+lookup table&0.000&0.018&0.000&0.020\\
  \noalign{\smallskip}\hline
\end{tabular}
\end{table}

\begin{table}
\centering
\begin{minipage}[]{90mm}
\caption[]{Statistics of the positional residuals for all stars after the analytical GD correction and the additional lookup table correction by using different orders polynomial fit. Column 1 is the date. Column 2 list the order of the polynomial fit. The following columns list the mean positional residual and its standard deviation (SD) in right ascension and declination, respectively. All units are in arcseconds.\label{tab4}}
\end{minipage}
\setlength{\tabcolsep}{1pt}
\small
 \begin{tabular}{ccccccccccccc}
 \\
  \hline\noalign{\smallskip}
  Date&  Orders          &$<$O-A$>$& SD & $<$O-A$>$& SD \\
      &                    &\multicolumn{2}{c}{RA}&\multicolumn{2}{c}{Dec.} \\
  \hline\noalign{\smallskip}
2016-01-17&2-order&0.000&0.035&0.000&0.039\\
          &3-order&0.000&0.018&0.000&0.021\\
          &4-order&0.000&0.018&0.000&0.021\\
          &5-order&0.000&0.018&0.000&0.021\\
2017-03-05&2-order&0.000&0.047&0.000&0.049\\
          &3-order&0.000&0.018&0.000&0.020\\
          &4-order&0.000&0.018&0.000&0.020\\
          &5-order&0.000&0.018&0.000&0.020\\
  \noalign{\smallskip}\hline
\end{tabular}
\end{table}

\section{Relative positions of the chips}
\label{sect:diss}
Due to the accurate GD pattern, it is meaningful to know the relative positions of the CCD array. As the center of the CCD array was adopted as the tangential point on the tangential plane, the standard coordinate for each pixel can be computed and the inter-chip gaps can be obtained. We use CCD$\#$2 and CCD$\#$3 as a reference to compute the inter-chip gap for each observational set (see Figure 2). As the inter-chip gaps were derived from the extrapolation of the final GD model, the far distance from the reference chip's boundary was not reliable for the measurement of the inter-chip gaps. As such, only twelve points were given to compute their positional changes. When CCD$\#$2 is used as the reference, the pixel positions for the corner 5, 6, and 8 are (0.5, 4031.5), (4031.5, 4031.5) and (4031.5, -0.5). When CCD$\#$3 is used as the reference, the pixel positions for the corner 9, 11, and 12 are (0.5, 4031.5), (0.5, -0.5) and (4031.5, -0.5). The pixel positions for the other corners are listed in Table~\ref{tab5}.

\begin{figure}
\centering
\includegraphics[width=8.5cm, angle=0]{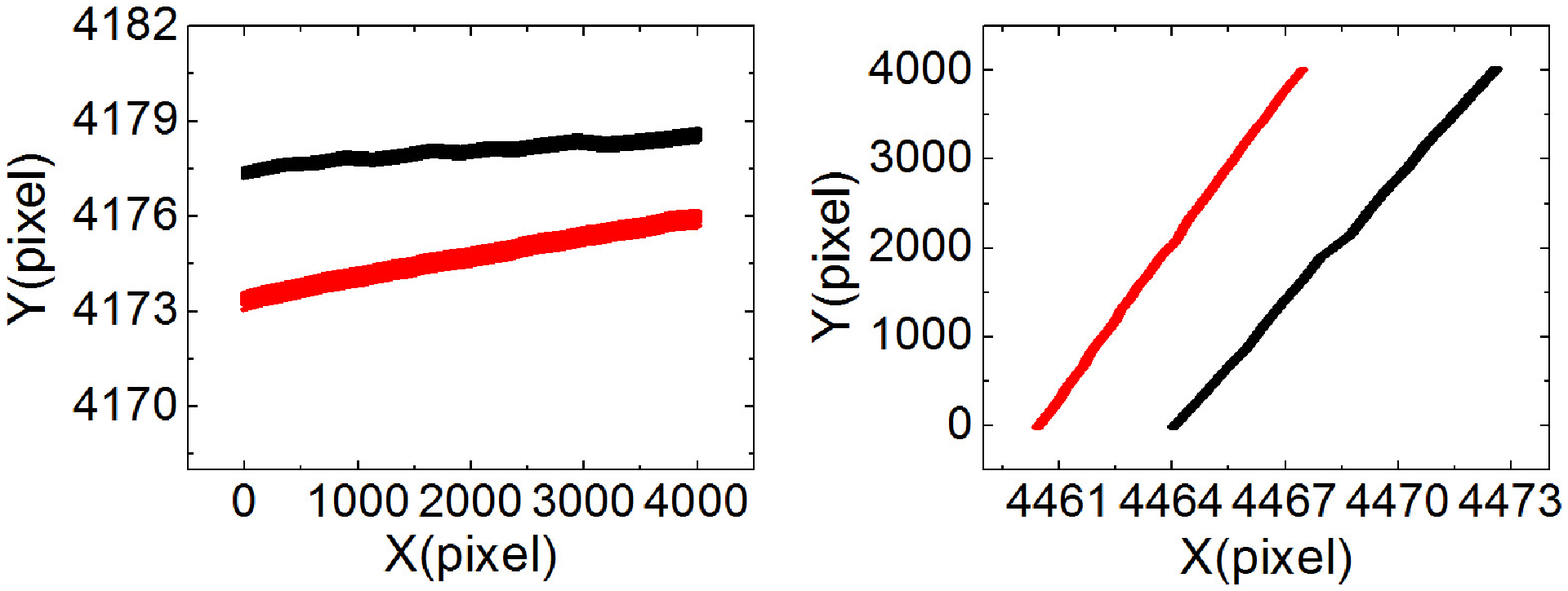}
\caption{The change of the inter-chip gap when CCD$\#$2 is used as the reference. The left panel shows the change in horizontal and the right panel shows the change in vertical. The black line shows the change in 2016 and the red line shows the change in 2017.}\label{Fig10_1}
\end{figure}

\begin{figure}
\centering
\centering
\includegraphics[width=8.5cm, angle=0]{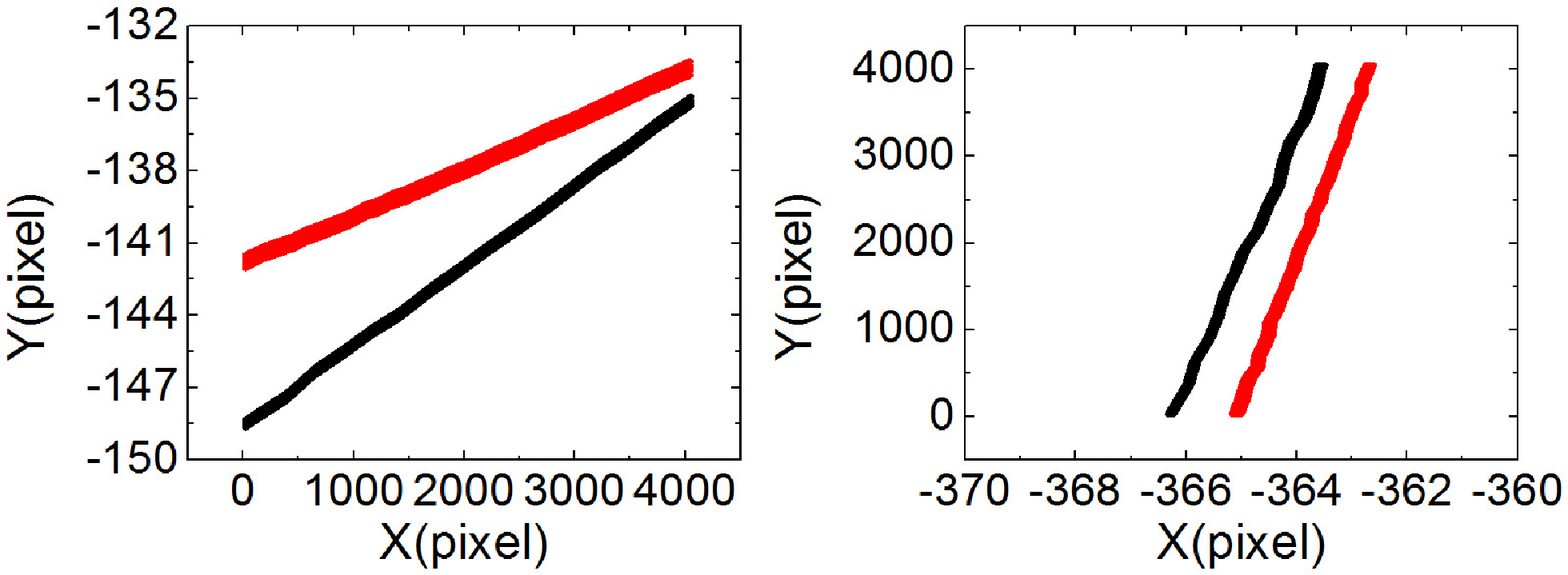}
\caption{The change of the inter-chip gap when CCD$\#$3 is used as the reference. The left panel shows the change in horizontal and the right panel shows the change in vertical. The black line shows the change in 2016 and the red line shows the change in 2017.}\label{Fig10_2}
\end{figure}

\begin{figure}
\centering
\includegraphics[width=8.5cm, angle=0]{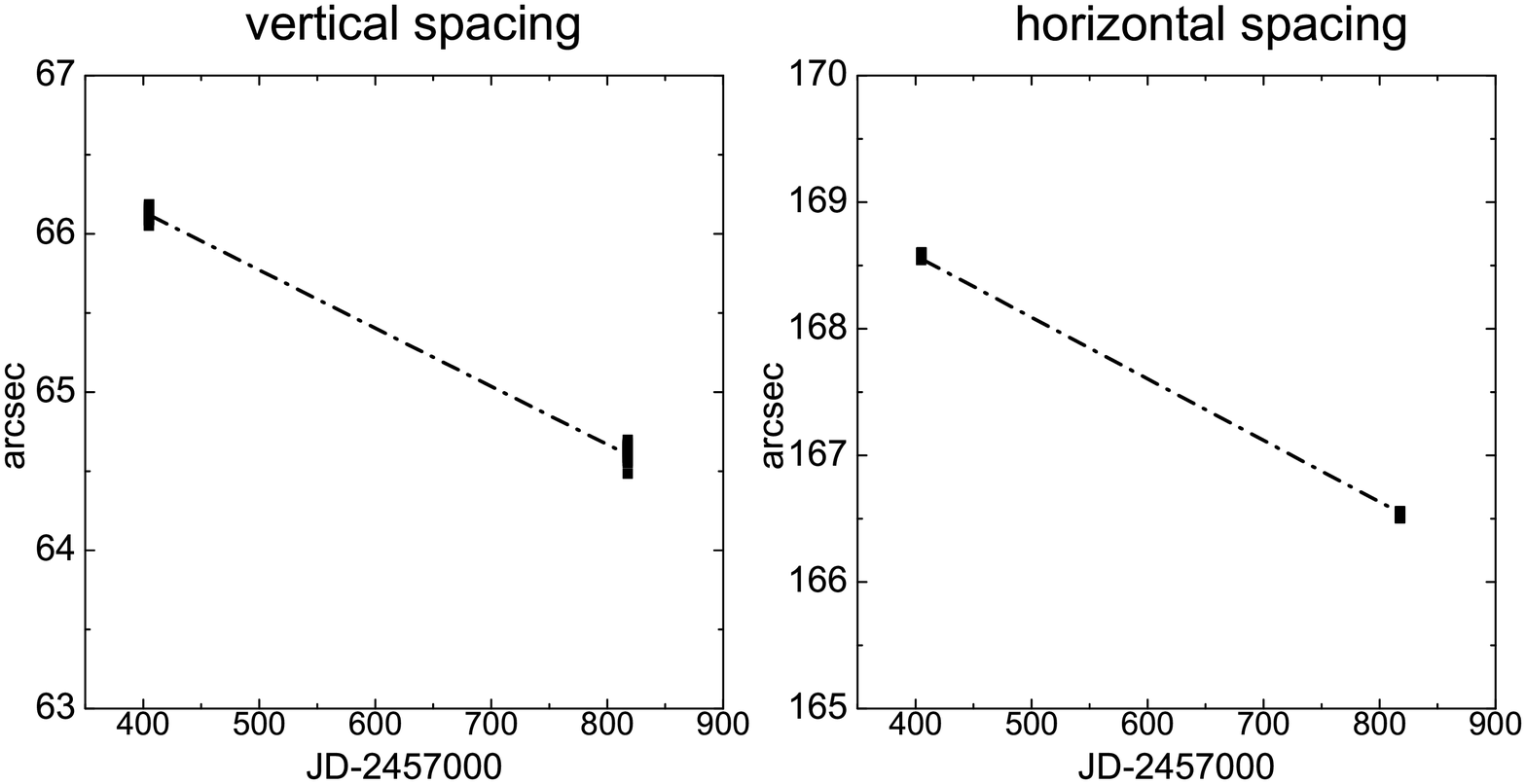}
\caption{The change of the inter-chip gap in 2016 and 2017. The left panel shows the change of the vertical gap with respect to CCD$\#$2, the right panel shows the change of the horizontal gap.}\label{Fig10_3}
\end{figure}

\begin{figure}
\centering
\centering
\includegraphics[width=8.5cm, angle=0]{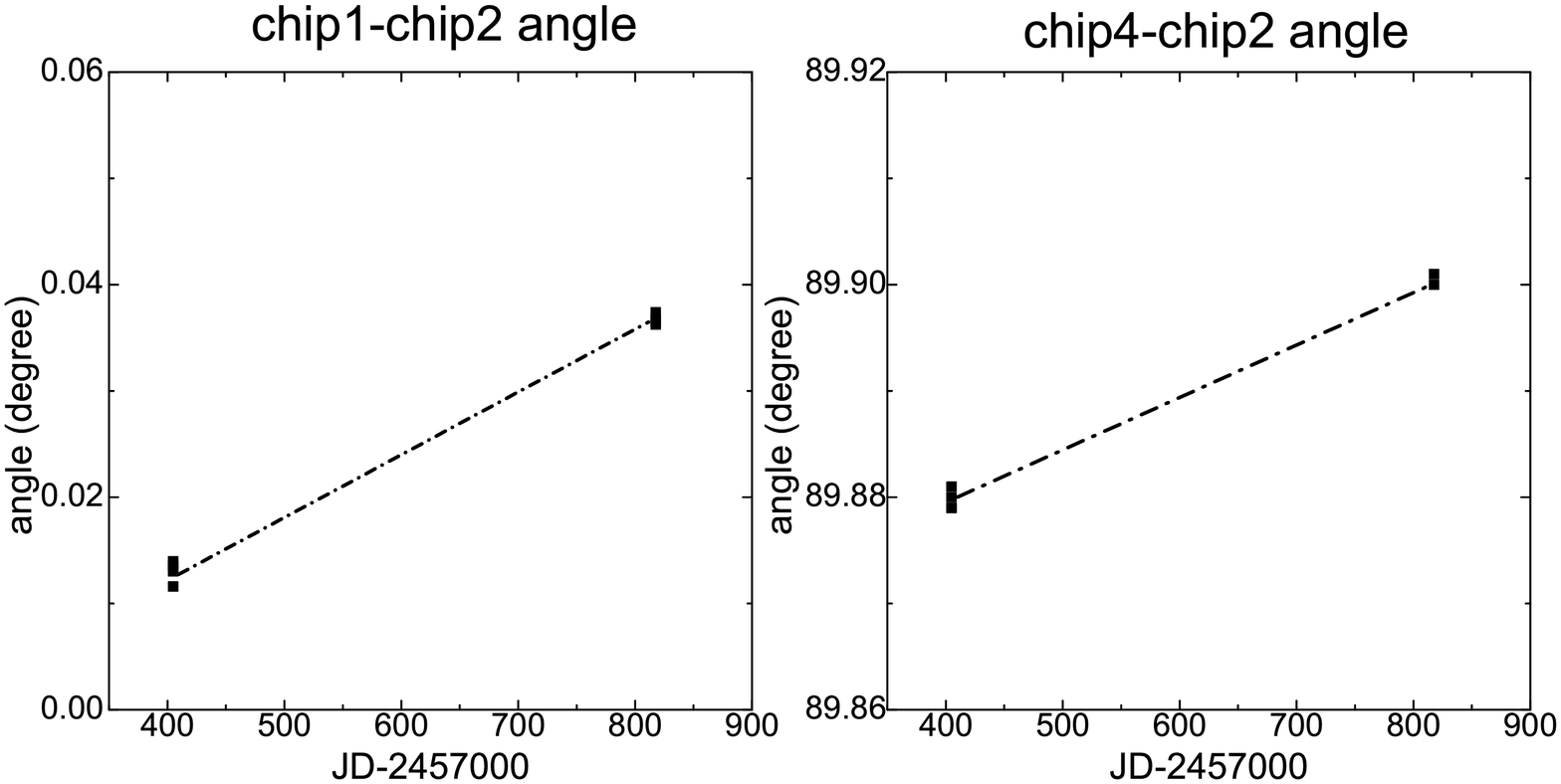}
\caption{The change of the roll angel in 2016 and 2017. The left panel shows the change of chip$\#$1 with respect to CCD$\#$2, the right panel shows the change of chip$\#$4 with respect to CCD$\#$2.}\label{Fig10_4}
\end{figure}

Figure~\ref{Fig10_1} shows the change of the inter-chip gap when CCD$\#$2 is used as the reference in 2016 and 2017, respectively, and Figure~\ref{Fig10_2} shows the change of the inter-chip gap when CCD$\#$3 is used as the reference. From Figure~\ref{Fig10_1} and Figure~\ref{Fig10_2} we can see there was a slight shift in the inter-chip gap in 14 months. Furthermore, Figure~\ref{Fig10_3} shows the change of the inter-chip gap and Figure~\ref{Fig10_4} shows the change of the roll angles between 2016 and 2017 with respect to CCD$\#$2 (chosen as reference). The statistics of inter-chip gaps and the roll angles in 2016 and 2017 are also shown in Table~\ref{tab6}. From Table~\ref{tab6} we can see that the change of the roll angle in 14 months is no more than 0.1 degree and the roll angles appear to be stable.

\begin{table*}
\centering
\begin{minipage}[]{180mm}
\caption[]{The positions of the points at the corners (see Figure 2). Column 2 to 9 show the mean pixel positions and their corresponding standard deviations. The top panel uses CCD$\#$2 as the reference and the bottom one CCD$\#$3 as the reference. All units are in pixels.\label{tab5}}
\end{minipage}
\setlength{\tabcolsep}{1pt}
\small
 \begin{tabular}{lccccccccc}
 \\
  \hline\noalign{\smallskip}
 Corner&\multicolumn{2}{c}{3}& \multicolumn{2}{c}{4}&\multicolumn{2}{c}{13}& \multicolumn{2}{c}{15} \\
 Date&  $<$mean$>$& SD & $<$mean$>$& SD&  $<$mean$>$& SD & $<$mean$>$& SD\\
  \hline\noalign{\smallskip}
2016&( -7.029,4177.380)&(0.063,0.059)&(4024.407,4178.564)&(0.069,0.089)&(4472.624,4012.683)&(0.042,0.242)&(4464.065,-17.983)&(0.034,0.059)\\
2017&(-10.275,4173.382)&(0.049,0.095)&(4020.632,4175.989)&(0.059,0.092)&(4467.475,4009.499)&(0.033,0.052)&(4460.441,-21.430)&(0.038,0.056)\\
  \noalign{\smallskip}\hline

\\
\hline
 Corner&\multicolumn{2}{c}{2}& \multicolumn{2}{c}{4}&\multicolumn{2}{c}{13}& \multicolumn{2}{c}{14} \\
 Date&  $<$mean$>$& SD & $<$mean$>$& SD&  $<$mean$>$& SD & $<$mean$>$& SD\\
  \hline\noalign{\smallskip}
2016&(-363.701,4053.247)&(0.039,0.067)&(-366.320,22.462)&(0.032,0.193)&(24.878,-148.552)&(0.063,0.073)&(4056.526,-135.165)&(0.051,0.082)\\
2017&(-362.698,4052.952)&(0.040,0.048)&(-365.073,22.179)&(0.042,0.060)&(20.084,-141.796)&(0.066,0.100)&(4050.305,-133.761)&(0.063,0.102)\\
  \noalign{\smallskip}\hline
\end{tabular}
\end{table*}

\begin{table}
\centering
\begin{minipage}[]{90mm}
\caption[]{Statistics of the inter-chip gaps and the roll angles. "16gap" is the abbreviation of the gap in 2016. "16angle" is the abbreviation of the angle in 2016. Column 2 to 9 show the averaged values and the standard deviation for the gaps and the angles. CCD$\#$2 and CCD$\#$3 are used as the reference. The unit for row 1 to 2 is in arcsec and for row 3 to 4 is in degree. The unit of the standard deviation for row 3 to 4 is in arcmin.\label{tab6}}
\end{minipage}
\setlength{\tabcolsep}{1pt}
\small
 \begin{tabular}{lccccccccc}
 \\
  \hline\noalign{\smallskip}
 Date&  $<$mean$>$& SD & $<$mean$>$& SD&  $<$mean$>$& SD & $<$mean$>$& SD\\
 Parameter&\multicolumn{2}{c}{CCD1(CCD2)}& \multicolumn{2}{c}{CCD4(CCD3)}&\multicolumn{2}{c}{CCD4(CCD2)}& \multicolumn{2}{c}{CCD1(CCD3)} \\
  \hline\noalign{\smallskip}
16gap&66.13&0.03&64.20&0.03&168.58&0.01&165.09&0.02\\
17gap&64.63&0.04&62.38&0.04&166.53&0.01&164.63&0.02\\
16angle&0.013&0.06&0.190&0.02&89.879&0.12&89.961&0.06\\
17angle&0.037&0.03&0.115&0.02&89.901&0.00&89.965&0.00\\
  \noalign{\smallskip}\hline
\end{tabular}
\end{table}

\section{Conclusion}
\label{sect:con}
Although the Bok telescope provides favourable astrometric properties, the GD must be solved carefully. The distortion solution for BASS on the Bok telescope is quite different from the previous method where the numerical GD model is enough. Instead of the numerical GD model, the analytical GD model and the additional lookup table were used to improve the astrometry of BASS. After the analytical GD correction and the additional lookup table correction are applied, the internal agreement or precision of the stars are estimated at about 20 mas and even better in each direction. This improvement can tap the astrometric potential of the Bok telescope and carry out some research work such as the reference frame, the brown dwarf, the parallax, the proper motion, and so on.

\section*{Acknowledgements}\emph{}
We acknowledge the support of the staff of the national Astronomical Observatory of China and the staff of the Bok telescope at Steward Observatory. This work was supported by the Joint Research Fund in Astronomy (U1431227) under cooperative agreement between the National Natural Science Foundation of China (NSFC) and Chinese Academy of Sciences (CAS), by the National Natural Science Foundation of China (Grant No. 11703008, 11873026), by the Natural Science Foundation of Guangdong Province, China (Grant No. 2016A030313092), by the Opening Project of Guangdong Province Key Laboratory of Computational Science at the Sun Yat-sen University, and partly by the Fundamental Research Funds for the Central Universities. This work has made use of data from the European Space Agency (ESA) mission {\it Gaia} (\url{https://www.cosmos.esa.int/gaia}), processed by
the {\it Gaia} Data Processing and Analysis Consortium (DPAC, \url{https://www.cosmos.esa.int/web/gaia/dpac/consortium}). Funding for the DPAC has been provided by national institutions, in particular the institutions participating in the {\it Gaia} Multilateral Agreement.


\bibliographystyle{mnras}
\bibliography{Bok} 

\begin{thebibliography}{}
\makeatletter
\relax
\def\mn@urlcharsother{\let\do\@makeother \do\$\do\&\do\#\do\^\do\_\do\%\do\~}
\def\mn@doi{\begingroup\mn@urlcharsother \@ifnextchar [ {\mn@doi@}
  {\mn@doi@[]}}
\def\mn@doi@[#1]#2{\def\@tempa{#1}\ifx\@tempa\@empty \href
  {http://dx.doi.org/#2} {doi:#2}\else \href {http://dx.doi.org/#2} {#1}\fi
  \endgroup}
\def\mn@eprint#1#2{\mn@eprint@#1:#2::\@nil}
\def\mn@eprint@arXiv#1{\href {http://arxiv.org/abs/#1} {{\tt arXiv:#1}}}
\def\mn@eprint@dblp#1{\href {http://dblp.uni-trier.de/rec/bibtex/#1.xml}
  {dblp:#1}}
\def\mn@eprint@#1:#2:#3:#4\@nil{\def\@tempa {#1}\def\@tempb {#2}\def\@tempc
  {#3}\ifx \@tempc \@empty \let \@tempc \@tempb \let \@tempb \@tempa \fi \ifx
  \@tempb \@empty \def\@tempb {arXiv}\fi \@ifundefined
  {mn@eprint@\@tempb}{\@tempb:\@tempc}{\expandafter \expandafter \csname
  mn@eprint@\@tempb\endcsname \expandafter{\@tempc}}}

\bibitem[\protect\citeauthoryear{{Anderson} \& {King}}{{Anderson} \&
  {King}}{2003}]{Anderson+2003}
{Anderson} J.,  {King} I.~R.,  2003, \mn@doi [\pasp] {10.1086/345491}, \href
  {http://adsabs.harvard.edu/abs/2003PASP..115..113A} {115, 113}

\bibitem[\protect\citeauthoryear{{Anderson}, {Bedin}, {Piotto}, {Yadav}  \&
  {Bellini}}{{Anderson} et~al.}{2006}]{Anderson+2006}
{Anderson} J.,  {Bedin} L.~R.,  {Piotto} G.,  {Yadav} R.~S.,   {Bellini} A.,
  2006, \mn@doi [\aap] {10.1051/0004-6361:20065004}, \href
  {http://adsabs.harvard.edu/abs/2006A%26A...454.1029A} {454, 1029}

\bibitem[\protect\citeauthoryear{{Bertin} \& {Arnouts}}{{Bertin} \&
  {Arnouts}}{1996}]{Bertin+1996}
{Bertin} E.,  {Arnouts} S.,  1996, \mn@doi [\aaps] {10.1051/aas:1996164}, \href
  {http://adsabs.harvard.edu/abs/1996A%26AS..117..393B} {117, 393}

\bibitem[\protect\citeauthoryear{{Blum} et~al.,}{{Blum} et~al.}{2016}]{blum+16}
{Blum} R.~D.,  et~al., 2016, in American Astronomical Society Meeting Abstracts
  \#228. p. 317.01

\bibitem[\protect\citeauthoryear{{Gaia Collaboration} et~al.,}{{Gaia
  Collaboration} et~al.}{2016a}]{gaia+16a}
{Gaia Collaboration} et~al., 2016a, \mn@doi [\aap]
  {10.1051/0004-6361/201629272}, \href
  {http://adsabs.harvard.edu/abs/2016A%26A...595A...1G} {595, A1}

\bibitem[\protect\citeauthoryear{{Gaia Collaboration} et~al.,}{{Gaia
  Collaboration} et~al.}{2016b}]{gaia+16b}
{Gaia Collaboration} et~al., 2016b, \mn@doi [\aap]
  {10.1051/0004-6361/201629512}, \href
  {http://adsabs.harvard.edu/abs/2016A%26A...595A...2G} {595, A2}

\bibitem[\protect\citeauthoryear{{Gaia Collaboration} et~al.,}{{Gaia
  Collaboration} et~al.}{2018}]{gaia+2018}
{Gaia Collaboration} et~al., 2018, \mn@doi [\aap]
  {10.1051/0004-6361/201833051}, \href
  {http://adsabs.harvard.edu/abs/2018A%26A...616A...1G} {616, A1}

\bibitem[\protect\citeauthoryear{Green}{Green}{1985}]{green1985}
Green R.~M.,  1985, Spherical Astronomy.
Cambridge Univ. Press, Cambridge

\bibitem[\protect\citeauthoryear{{Li}, {Peng}  \& {Han}}{{Li}
  et~al.}{2009}]{liz+09}
{Li} Z.,  {Peng} Q.~Y.,   {Han} G.~Q.,  2009, Acta Astronomica Sinica, \href
  {http://adsabs.harvard.edu/abs/2009AcASn..50..340L} {50, 340}

\bibitem[\protect\citeauthoryear{{Peng}, {Vienne}, {Zhang}, {Desmars}, {Yang}
  \& {He}}{{Peng} et~al.}{2012}]{Peng+etal+2012}
{Peng} Q.~Y.,  {Vienne} A.,  {Zhang} Q.~F.,  {Desmars} J.,  {Yang} C.~Y.,
  {He} H.~F.,  2012, \mn@doi [\aj] {10.1088/0004-6256/144/6/170}, \href
  {http://adsabs.harvard.edu/abs/2012AJ....144..170P} {144, 170}

\bibitem[\protect\citeauthoryear{{Peng}, {Wang}, {Vienne}, {Zhang}, {Li}  \&
  {Meng}}{{Peng} et~al.}{2015}]{Peng+etal+2015}
{Peng} Q.~Y.,  {Wang} N.,  {Vienne} A.,  {Zhang} Q.~F.,  {Li} Z.,   {Meng}
  X.~H.,  2015, \mn@doi [\mnras] {10.1093/mnras/stv469}, \href
  {http://adsabs.harvard.edu/abs/2015MNRAS.449.2638P} {449, 2638}

\bibitem[\protect\citeauthoryear{{Peng}, {Peng}  \& {Wang}}{{Peng}
  et~al.}{2017}]{Peng+etal+2017}
{Peng} H.~W.,  {Peng} Q.~Y.,   {Wang} N.,  2017, \mn@doi [\mnras]
  {10.1093/mnras/stx229}, \href
  {http://adsabs.harvard.edu/abs/2017MNRAS.467.2266P} {467, 2266}

\bibitem[\protect\citeauthoryear{{Ren} \& {Peng}}{{Ren} \&
  {Peng}}{2017}]{ren+2010}
{Ren} J.~J.,  {Peng} Q.~Y.,  2017, Astronomical research technology, 7, 115

\bibitem[\protect\citeauthoryear{{Silva} et~al.,}{{Silva}
  et~al.}{2016}]{silva+16}
{Silva} D.~R.,  et~al., 2016, in American Astronomical Society Meeting
  Abstracts \#228. p. 317.02

\bibitem[\protect\citeauthoryear{{Wang}, {Peng}, {Zhang}, {Zhang}, {Li}  \&
  {Meng}}{{Wang} et~al.}{2015}]{Wang+etal+2015}
{Wang} N.,  {Peng} Q.~Y.,  {Zhang} X.~L.,  {Zhang} Q.~F.,  {Li} Z.,   {Meng}
  X.~H.,  2015, \mn@doi [\mnras] {10.1093/mnras/stv2236}, \href
  {http://adsabs.harvard.edu/abs/2015MNRAS.454.3805W} {454, 3805}

\bibitem[\protect\citeauthoryear{{Wang}, {Peng}, {Peng}, {Xie}, {Ma}  \&
  {Zhang}}{{Wang} et~al.}{2017}]{Wang+etal+2017}
{Wang} N.,  {Peng} Q.~Y.,  {Peng} H.~W.,  {Xie} H.~J.,  {Ma} S.,   {Zhang}
  Q.~F.,  2017, \mn@doi [\mnras] {10.1093/mnras/stx550}, \href
  {http://adsabs.harvard.edu/abs/2017MNRAS.468.1415W} {468, 1415}

\bibitem[\protect\citeauthoryear{{Zhou} et~al.,}{{Zhou} et~al.}{2016}]{zhou+16}
{Zhou} X.,  et~al., 2016, \mn@doi [Research in Astronomy and Astrophysics]
  {10.1088/1674-4527/16/4/069}, \href
  {http://adsabs.harvard.edu/abs/2016RAA....16...69Z} {16, 69}

\bibitem[\protect\citeauthoryear{{Zou} et~al.,}{{Zou}
  et~al.}{2017}]{zou+etal+2017+aj}
{Zou} H.,  et~al., 2017, \mn@doi [\aj] {10.3847/1538-3881/aa72d9}, \href
  {http://adsabs.harvard.edu/abs/2017AJ....153..276Z} {153, 276}

\makeatother
\end{thebibliography}

\appendix

\section{The header of the Bok's fits file}
Here, Table~\ref{tab7} lists a part of a typical FITS header in this work.
\begin{table}
\centering
\begin{minipage}[]{90mm}
\caption[]{The part of the FITS header. The first column lists the keywords. Column 3 lists the value and column 4 lists the description for the keyword and the value.\label{tab7}}
\end{minipage}
\setlength{\tabcolsep}{1pt}
\small
 \begin{tabular}{llrl} \\
SIMPLE  &=     &           T   & conforms to FITS standard\\
BITPIX  &=     &           -32 & array data type\\
NAXIS   &=     &           2   & number of array dimensions\\
NAXIS1  &=     &           4096& \\
NAXIS2  &=     &           4032& \\
RA-OBS  &=     &  '10:46:11.97'& right ascension (2000.)\\
DEC-OBS &=     &  '+28:15:47.1'& declination\\
DATE-OBS&=     &'2016-01-17T10:24:38.506'&UTC shutter opened\\
TIME-OBS&=     &  '10:24:38.506'&UTC at start of exposure\\
TIMESYS &=     &           UTC& Time system\\
EXPTIME &=     &          60.0& Exposure time (seconds)\\
HA      &=     &    '-00:03:50'&hour angle \\
ST      &=     &    '10:43:15' &local siderial time\\
FILTER  &=     &     'bokr    '&Filter name\\
FOCUSVAL&=     &'*0.761*0.654*0.703'&Focus\\
INSTRUME&=     &    '90prime ' &Instrument name\\
SITEELEV&=     &     '2120    '&\\
SITELAT &=     &      '31:58.8'&\\
SITELONG&=     &    '-111:36.0'&\\
CAMTEMP &=     &       -134.609&Camera temperature in C\\
DEWTEMP &=     &       -191.452&Dewar temperature in C\\
CRPIX1  &=     &  -1.820100000000E+02&Azimuth=167.9\\
CRPIX2  &=     &  -5.904000000000E+01&Altitude=86.11\\
AIRMASS &=     &       1.002   &\\
MJD-OBS &=     &    57404.43378&\\
RADESYS &=     &     'ICRS    '&Astrometric system\\
\end{tabular}
\end{table}

\bsp	
\label{lastpage}

\end{document}